\documentclass[a4paper]{article}

\usepackage{INTERSPEECH2021}
\usepackage[bottom]{footmisc}
\usepackage{epsfig,amssymb,amsmath,rotating}
\ninept
\usepackage{pgfplots,balance}
\usepackage{grffile}
\usepackage{float}
\PassOptionsToPackage{hyphens}{url}
\usepackage[bookmarks=false,hidelinks]{hyperref}

\usepackage[caption=false,font=footnotesize,subrefformat=parens,labelformat=parens]{subfig}
\usepackage{tikz}
\usepackage{pgfplots}
\usepackage{grffile}
\usepackage{float}
\pgfplotsset{compat=newest}
\makeatletter
\g@addto@macro\@floatboxreset\centering
\makeatother
\def\addlegendimage{\csname pgfplots@addlegendimage\endcsname}
\usepackage{graphicx}
\usepackage{cmap}
\usetikzlibrary{shapes.misc}
\usetikzlibrary{mindmap}
\usetikzlibrary{plotmarks}
\usetikzlibrary{arrows.meta}
\usepgfplotslibrary{patchplots}

\tikzset{
    font=\scriptsize
}

\newcommand{\setvariable}[2]{
	\let#1\relax
	\newcommand{#1}{#2}
}

\renewcommand{\vec}[1]{\bm{#1}}

\definecolor{darkgreen}{rgb}{0.2,0.7,0.3}
\definecolor{darkorange}{rgb}{1,0.55,0}

\pgfplotsset{
    mated/.style={ybar interval, fill opacity=0.6, area legend, 
        fill=darkgreen, draw=white},
    nonmated/.style={ybar interval, fill opacity=0.6, area legend, 
        fill=white!30!red, draw=white},
    histAxis/.style={
        width=\figwidth,
        height=\figheight,
        scale only axis,
        ymin=0,ymax=1,
        ylabel={Rel. freq.},
        xmajorgrids,ymajorgrids,},
    rawHistAxis/.style={histAxis,xlabel={Scores},xmin=-20,xmax=40},
    llrHistAxis/.style={histAxis,xlabel={Ideal LLRs},xmin=-10,xmax=10},
}

\usepackage{eurosym}

\setvariable{\colorScalePurple}{PiYG-11-1}
\setvariable{\colorScaleOrange}{Oranges-9-4}
\setvariable{\colorScaleBlue}{Spectral-5-5}
\setvariable{\colorScaleRed}{hda}
\setvariable{\colorScaleGreen}{Spectral-5-4!80!black}

\usepackage{amsmath}
\usepackage{amsfonts}
\usepackage{amssymb}
\usepackage{url}
\usepackage{graphicx}
\usepackage{epstopdf}
\usepackage{mwe}
\usepackage{multirow}
\usepackage[export]{adjustbox}
\usepackage{tikz}
\usepackage{algorithm}
\usepackage{kantlipsum}
\usepackage{xpatch,xcolor}
\usepackage{graphicx}
\usepackage{afterpage}
\usepackage{balance}
\usepackage{cite}

\usepackage{xcolor}
\makeatletter
\usepackage{tabularx}
\usepackage{multirow}

\makeatother
\tikzset{
    font=\scriptsize
}
\hypersetup{
    colorlinks=true,
    linkcolor=purple,
    filecolor=magenta,      
    urlcolor=purple,
}
\newcolumntype{Y}{>{\centering\arraybackslash}X}

\usepackage{balance}

\title{Graph Attention Networks for Anti-Spoofing}
\name{Hemlata Tak$^1$, Jee-weon Jung$^2$, Jose Patino$^1$, Massimiliano Todisco$^1$ and Nicholas Evans$^1$}

\address{
  $^1$EURECOM, Sophia Antipolis, France\\
  $^2$Naver Corporation, South Korea}
\email{lastname@eurecom.fr, jeeweon.jung@navercorp.com}

\begin{document}

\maketitle

\begin{abstract}
  The cues needed to detect spoofing attacks against automatic speaker verification are often located in specific spectral sub-bands or temporal segments.
  Previous works show the potential to learn these using either spectral or temporal self-attention mechanisms but not the relationships between neighbouring sub-bands or segments.
  This paper reports our use of graph attention networks (GATs) to model these relationships and to improve spoofing detection performance. 
  GATs leverage a self-attention mechanism over graph structured data to model the data manifold and the relationships between nodes. 
  Our graph is constructed from representations produced by a ResNet. 
  Nodes in the graph represent information either in specific sub-bands or temporal segments.
  Experiments performed on the ASVspoof 2019 logical access database show that our GAT-based model with temporal attention outperforms all of our baseline single systems. 
  Furthermore, GAT-based systems are complementary to a set of existing systems. The fusion of GAT-based models with more conventional countermeasures delivers a 47\% relative improvement in performance compared to the best performing single GAT system.
\end{abstract}

\vspace{0.2cm}


\section{Introduction}
The success of anti-spoofing solutions for automatic speaker verification (ASV) systems is dependent on the reliable identification of processing artefacts stemming from the manipulation or synthesis of speech signals~\cite{nautsch2021asvspoof}. 
These artefacts are known to reside within specific sub-bands or temporal segments~\cite{ sriskandaraja2016investigation,yang2019significance,garg2019subband,tomi2020subband,odyssey2020CQCC,IS2020LFCC}. 
Their detection hence calls for spoofing countermeasure systems with spectral and/or temporal attention.

Convolutional neural network (CNN) approaches have been applied extensively to the anti-spoofing problem. 
CNNs are particularly appealing because of their capacity to extract localised artefacts within spectro-temporal decompositions such as a spectrogram.
For both the ASVspoof 2017~\cite{lavrentyeva2017audio} and ASVspoof 2019~\cite{lavrentyeva2019stc,chettri2019ensemble,lei2020siamese} challenges, CNN-based approaches were among the best performing systems. More elaborate systems, such as those based upon ResNet architectures, are now attracting greater interest, and enable the learning of deeper networks using residual blocks with skip connections~\cite{lai2019assert,alzantot2019deep,parasu2020investigating,aravind2020audio,zhang2020one,chen2020generalization,li2020replay,wang2021comparative}.

Our own work~\cite{tak2021rawnet} explored the use of a RawNet2 architecture~\cite{jung2020improved}.
Similar to the original RawNet architecture~\cite{jung2019rawnet}, RawNet2 adopts residual blocks with skip connections and feeds their output to a gated recurrent unit layer to extract utterance level representations. The first convolutional layer uses a bank of band-pass filters parametrised in the form of sinc functions in identical fashion to SincNet~\cite{ravanelli2018speaker}. 
Other differences include the use of cosine similarity scoring and the application of feature map scaling (FMS) to residual block outputs.  
FMS acts to emphasise the most salient sinc filter outputs (or sub-bands) by applying different weights to aggregated features extracted from each band.
The network can hence be optimised to apply greater attention to the most discriminant sub-bands. FMS can be interpreted as a form of spectral attention and is applied to residual block outputs which is an aggregated frequency response. However, it neither learns nor models the relationships between different filters or sub-bands (e.g.\ spoofing artefacts present simultaneously in two different sub-bands). Like CNNs, ResNet and RawNet-based architectures also lack the capacity to capture and use such information to help discriminate between bona fide and spoofed speech.
Having observed that spoofing artefacts can be present across multiple sub-bands or temporal segments, our hypothesis is that an attention mechanism with the power to model the relationships between different them has potential to improve upon anti-spoofing performance.

Graph neural networks (GNNs)~\cite{gori2005new,scarselli2008graph}, especially recent architectures such as graph convolution networks (GCNs)~\cite{Kipf2017Semi-supervisedNetworks} or graph attention networks (GATs)~\cite{velivckovic2017graph} can be used to model these relationships.
Instead of modelling frames or sub-band representations linearly, GNNs models the non-Euclidean data manifold spanning different sub-bands and temporal segments.
While conventional attention mechanisms can be used in spatial or sequence-based tasks to focus on more relevant information, graph attention mechanisms have the capacity to learn which sub-bands or segments are the most informative with regard to their neighbours, and to assign weights to emphasise those that are the most discriminative.
Accordingly, we have explored the use of GNN-based architectures to model spectral and temporal relationships for the spoofing detection task.

The remainder of the paper is organised as follows.  Section~2 describes the related work, Section~3 introduces the proposed GAT approach to anti-spoofing. The experimental setup and results are presented in Section~4. Finally, the paper is concluded in Section~5.

\begin{figure*}[t]
  \centering
  \includegraphics[width=\linewidth]{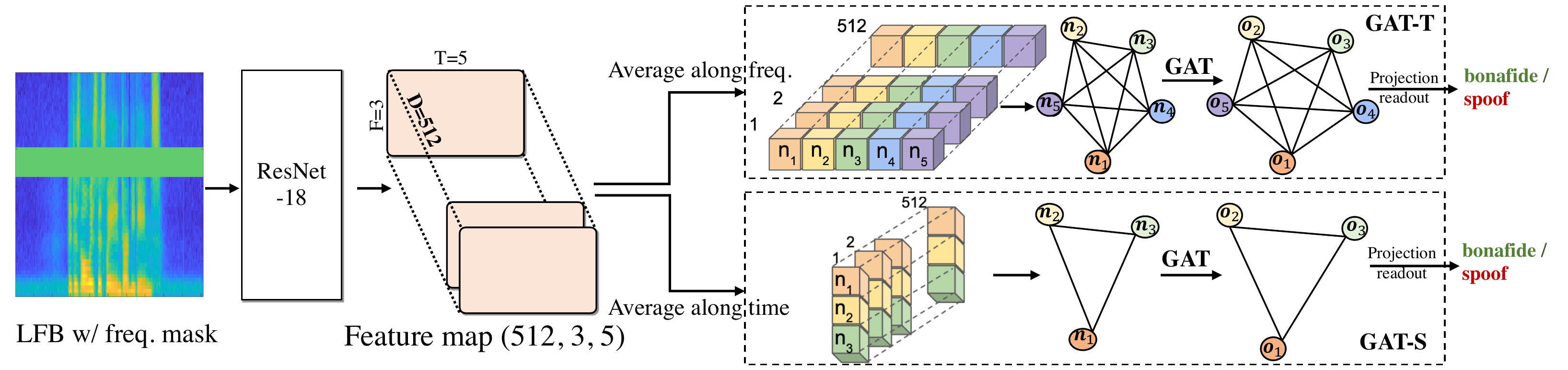}
  \caption{
    Proposed GAT approach to anti-spoofing.
    Frequency-masked, high-resolution log-linear filterbank (LFB) features extracted from an input utterance are fed to a ResNet-18 network to extract high-level representations. 
    The ResNet-18 feature map output is used to construct a graph with nodes $n$ after averaging across time, (spectral attention, \textbf{GAT-S}) or frequency (temporal attention, \textbf{GAT-T}). 
    We use a GAT layer followed by projection and readout to predict whether the input utterance is bona fide or spoofed.}
   
  \label{fig:GAT pipeline}
\end{figure*}

\vspace{0.4cm}

\section{Related work}

Velickovic et al.~\cite{velivckovic2017graph} introduced graph attention networks (GATs) for node classification in graph-structured data. 
By modelling the relationships between neighbouring nodes using a self-attention approach, GAT was used to learn the hidden representations of each node in the graph. The work demonstrated that GAT-based attention models can also be applied to arbitrarily structured graphs. Rather than modelling these relationships with identical weights as with GraphSage~\cite{hamilton2017inductive} or with pre-determined weights as with GCNs~\cite{Kipf2017Semi-supervisedNetworks}, GATs learn weights via a self-attention mechanism, such that nodes are weighted according to the information they provide relative to neighbouring nodes~\cite{velivckovic2017graph}.

GNNs, including GCN and GAT variants, have been applied successfully in a variety of speech-related tasks. 
Zhang et al.~\cite{zhang2019few} applied GCNs to few-shot audio classification.
GNNs were used to predict an attention vector which helps to discriminate between different audio classes accordingly to their relative importance. Results showed the effectiveness of so-called attentional GNNs in transferring the metric representation learned from training classes to novel classes. 
Liu et al.~\cite{liu2020graphspeech} demonstrated the application of GNNs to neural speech synthesis, for which they are used to encode explicitly the syntactic relationship of the different elements within a sentence. Jung et al.~\cite{jung2020graph} showed how GATs can be used to learn utterance-level relationships between speakers and how a GAT architecture with residual connections can be adapted to compute utterance-level similarity scores for speaker verification. 
To the best of our knowledge, this paper reports the first application of GATs to the anti-spoofing problem for which it is used to model the relationships between isolated spectral or temporal artefacts.

\section{Anti-spoofing using GATs}

In this section, we describe our approach to anti-spoofing using GATs.  We describe the deep residual network (ResNet-18) architecture~\cite{he2016deep} which is used to extract high-level representation and then the GAT framework.  Finally, we discuss the application of temporal and spectral attention.  The architecture is illustrated in Figure~\ref{fig:GAT pipeline}.

\subsection{Extraction of high-level representations}

\begin{table}[!t]

	\caption{The details of ResNet-18 architecture. Convolutional layers are followed by batch normalisation and scaled exponential linear unit (selu) activation function. The output from average layer is use for graph formulation for GAT where averaging is applied along time or frequency domain depending on the configuration. Numbers denoted in output size are refers to (no.\ of CNN filters $\times$ frequency $\times$ time).}
	\centering
  \setlength\tabcolsep{3.2pt}
	\begin{tabularx}{\linewidth}{l|c|c|c|Y}
		\hline

		 Layer & Kernel & Filters&Stride&Output size	\\ 

		\hline
	Freq. masking&-&-&-&60 $\times$202\\
	Convolutional &3$\times$3&64&1$\times$2&$64\times$64$\times$103\\
	Max pooling&3$\times$3&-&2$\times$2&64$\times$32$\times$52\\
	Res. block (1)&3$\times$3&64&2$\times$2&64$\times$32$\times$52\\
	Res. block (2)&3$\times$3&128&2$\times$2&128$\times$16$\times$26\\
	Res. block (3)&3$\times$3&256&2$\times$2&256$\times$8$\times$13\\
	Res. block (4)&3$\times$3&512&2$\times$2&512$\times$3$\times$5\\
	Avg. along freq. &-&-&-&512$\times$5\\
	Avg. along time &-&-&-&512$\times$3\\
	
	\hline
	\end{tabularx}
	\label{Tab:ResNet18}
	\vspace{-0.15cm}
\end{table}

We use a ResNet-18~\cite{he2016deep} system to learn high-level representations from acoustic features. The use of residual blocks facilitates the learning of a deeper network than is possible without them. The architecture is illustrated in Table~\ref{Tab:ResNet18}.  The network consists of a 3$\times$3 convolutional layer and a 3$\times$3 max pool layer to downsample the input feature map. Then follows four residual blocks and averaging either in time or frequency, depending on the domain in which attention is applied.
The kernel, filter, stride and output sizes (no.\ of CNN filters $\times$ frequency $\times$ time) for each layer are illustrated in Table~\ref{Tab:ResNet18}.

\subsection{Graph Attention Network}
\label{section:GAT}

High-level feature representations are fed to the GAT whose architecture is identical to that in~\cite{jung2020graph}.
A graph $\mathcal{G}$ is first formed from the ResNet-18 output $e$ where $e\in\mathbb{R}^{N \times D}$, where $N$ is the number of nodes ($N$=3 or 5 depending on whether the ResNet-18 output is averaged across time or frequency) and where $D$=512 is the feature/node dimension.
The graph is \textit{fully-connected} with edges between every pair of nodes, including self-connections.
A GAT layer aggregates neighboring nodes using weights learned with a self-attention mechanism. Through this process, nodes are projected into another representation learned from the minimisation of a training loss. Using a GAT, more informative nodes are aggregated using greater weights, where the weight reflects the strength of the relationship between a given node pair.\\

The GAT output is denoted as:
\begin{equation}
    GAT(\mathcal{G})=\frac{1}{N}\sum_{n \in \mathcal{G}}\boldsymbol{o}_{n}W_{out},
\end{equation}
where $W_{out}$ is the projection matrix which maps each node vector to a scalar and $o_n$ is the output feature for node $n$ which is determined according to GAT node propagation:
\begin{equation}
    o_n=BN(W_{att}(\boldsymbol{m}_{n}) + W_{res}(\boldsymbol{e}_{n})),
\end{equation}
where BN is batch normalisation~\cite{Ioffe2015BatchShift}, $W_{att}$ is a matrix which projects the aggregated information for each node $n$ to the target dimensionality, whereas $W_{res}$ projects the residual to match the target dimensionality.

The information from neighboring nodes is aggregated via self-attention according to:
\begin{equation}
    m_n=\sum_{v \in \mathcal{M}(n) \cup \{ n \} }\alpha_{v,n}e_{v},
    \label{Eq:node aggregation}
\end{equation}
where $\mathcal{M}(n)$ refers to the neighbouring nodes of node $n$, and $\alpha_{v,n}$ refers to the attention weight between nodes $v$ and $n$. We consider the neighbouring nodes for node $n$ to be the full set of nodes within the graph, including  the node itself.
The attention weight is calculated differently that in the original work~\cite{velivckovic2017graph} according to:

\begin{equation}
    \alpha_{v,n}=\frac{{\operatorname{exp}}(W_{map}(e_{n} \odot e_{v}))}
    {\sum_{w \in \mathcal{M}(n) \cup \{ n \}}  {\operatorname{exp}}(W_{map}(e_{n} \odot e_{w}))  },
\end{equation}
where $W_{map} \in \mathcal{R}^{D}$ is the learnable map applied to the dot product and where $\odot$ denotes element-wise multiplication. Full details are available in~\cite{jung2020graph}.

\subsection{Spectral and temporal attention}
Psychoacoustics research~\cite{Brown91} shows that the human auditory system can select the most informative spectral bands and acts to perform an auto-correlation corresponding to the temporal correlation between adjacent frames. In order to capture such cues, we apply GATs with attention in either spectral or temporal domains. Temporal attention (GAT-T, top-right of Figure~\ref{fig:GAT pipeline}) is applied to model the temporal relationships between adjacent frames and can help to capture complex nonlinear temporal artefacts.
Spectral attention (GAT-S, bottom-right of Figure~\ref{fig:GAT pipeline}) is used to model the relationships between different sub-bands.

\section{Experimental setup}
Our work was performed using the ASVspoof 2019 Logical Access (LA) database~\cite{wang2020asvspoof} and default metrics.  We report results for our specific implementation of GAT solutions with either temporal or spectral attention and compare these to results for competing, state-of-the-art systems.

\subsection{Database and evaluation metric}

The ASVspoof 2019 LA database has three independent subsets: train; development; evaluation. 
Spoofed speech in each dataset is generated using a set of different  
speech synthesis, voice conversion and hybrid algorithms~\cite{wang2020asvspoof}. Attacks in the training and development set were created with a set of 6 different algorithms (A01-A06), whereas those in the evaluation set were created with a set of 13 algorithms (A07-A19).  We used the minimum normalised tandem detection cost function (t-DCF)~\cite{kinnunen2018t} as a primary metric but also report results in terms of the pooled equal error rate (EER).

\subsection{Baselines}
We implemented three baselines: a high-spectral resolution linear frequency cepstral coefficient  system with a conventional Gaussian mixture model classifier~\cite{IS2020LFCC} (LFCC-GMM); a ResNet-18 system (the same as used in our GAT system, but with different attention mechanisms); a RawNet2 system~\cite{tak2021rawnet}. 
We used the ResNet-18 systems in order to compare the benefit of GAT-based attention to alternative attention mechanisms: statistics pooling (SP)~\cite{snyder2017deep}; self-attentive pooling (SAP) which assigns different weights to different frames using a weighted mean; attentive statistical pooling (ASP)~\cite{okabe2018attentive} which generates different weights for different frames according to both weighted means and weighted standard deviations.

\begin{table*}[!ht]
\def\arraystretch{1.1}
	\small
	\centering
	
	\caption{Results for the ASVspoof 2019 logical access (LA) database in terms of min t-DCF for each attack in the development (A01-A06) and evaluation (A07-A19) partitions.  Pooled min~t-DCF (P1) and pooled EER (P2) are also shown for each partition.  Results shown for the baseline systems and the proposed GAT systems with temporal attention (GAT-T) and spectral attention (GAT-S).}

	\setlength\tabcolsep{1.5pt}
	\begin{tabularx}{\textwidth}{l || *{6}{Y} |YY| *{13}{Y}| YY} 
		\hline
		
		System &A01&A02&A03&A04&A05&A06&P1&P2&A07&A08&A09&A10&A11&A12&A13&A14&A15&A16&A17&A18 &A19& P1&P2 	\\ 
		\hline\hline

		LFCC-GMM&.000&.000&.000&.001&.000&.000&.000&.00&.001&.001&.000&.154&.005&.115&.080&.069&.069&.006&.352&.074&.008&.090 &	\textbf{3.50}\\
		
		RawNet2&.030&.020&.015&.043&.036&.042&.036&1.1&.098&.179&.073&.089&.042&.088&.020&.013&.073&.046&.240&.629&.058&.155&5.54\\
	    
	    GAT-T&.000&.000&.000&.000&.000&.000&.000&.00&.000&.009&.001&.012&.009&.011&.016&.010&.009&.000&.715&.073&.001&\textbf{.089}&4.71\\
		
		GAT-S&.000&.000&.000&.000&.000&.000&.000&.00&.000&.012&.000&.009&.006&.006&.009&.008&.009&.000&.642&.088&.000&.091&4.48\\
		
		ResNet18-SP&.001&.000&.000&.002&.000&.002&.002&.07&.001&.025&.001&.009&.009&.006&.011&.010&.010&.001&.966&.183&.005&.114&6.82\\
		
		ResNet18-SAP&.001&.000&.000&.001&.001&.000&.000&.03&.002&.080&.011&.023&.015&.029&.053&.037&.040&.001&.944&.272&.005&.138&7.11\\
		
		ResNet18-ASP&.000&.000&.000&.001&.000&.000&.000&.01&.000&.037&.001&.009&.008&.007&.010&.010&.009&.001&.809&.291&.006&.127&6.22\\

		\hline
	\end{tabularx}
	\label{Tab: results on dev and eval}
\end{table*}

\subsection{Implementation details} 

While we obtained similar results using alternative representations such as linear frequency cepstral coefficients (LFCCs)~\cite{sahidullah2015comparison,IS2020LFCC}, all work reported in this paper was performed with $60$-dimensional log linear filter bank (LFB) features extracted from 30~ms windows with a 10~ms frame shift and from audio waveforms which are truncated or concatenated to {$\approx $}{4} second segments (64600 samples). 
To improve generalisation, we applied frequency masking augmentation~\cite{park2019specaugment,chen2020generalization} to mask a random selection of contiguous frequency bands during training. The same frequency mask is applied to all training data within the same mini-batch. 
The maximum number of masked frequency bands was set to $12$.

The ResNet-18 system operates upon the LFB features to produce high-level feature representations which are fed to the GAT after either temporal or spectral averaging.  These are used to form an input graph where each node represents sub-bands or temporal segments with a 512-dimensional nodes/features.  
The GAT operates upon the input graph to generate the weighted output graph where the  nodes/features dimension is $128$. 
The latter are projected into a one-dimensional space (scalar) using a affine transforms (i.e.\ dense layer). 
Node features are finally aggregated using a readout layer to predict the output score. The entire system (both ResNet-18 and GAT) is trained using the ASVspoof 2019 LA training partition with binary cross-entropy (BCE) loss with a sigmoid activation function, a fixed learning rate of $0.0001$ and a weight decay parameter of $0.0001$. 
We used the standard Adam optimiser with a mini-batch size of $64$ and train for $300$ epochs.

\subsection{Results}

Results in terms of the min t-DCF are illustrated in Table~\ref{Tab: results on dev and eval} for development data (A01-06) and evaluation data (A07-A19).  Columns labelled P1 and P2 show pooled min t-DCF and pooled EER results for each partition. For the development set, the baseline LFCC-GMM, GAT-T, GAT-S and ResNet-18 systems all yield min t-DCF values of close to zero, whereas the RawNet-2 system compares poorly. For almost all attacks in the evaluation set, GAT-T and GAT-S systems perform as well as, or better than the baseline systems. While pooled results show only a modest improvement for the GAT-T system, they are dominated by results for the A17 attack, for which the best results are obtained by the RawNet2 system. A comparison of results for other attacks and for ResNet-18, GAT-T and GAT-S systems show the benefit of graph-based approaches to model temporal or spectral relationships.  The attention mechanisms of the three ResNet-18 systems are less effective, with pooled results being worse than for the LFCC-GMM system without any attention mechanism. We also note substantial differences in performance at the attack level.  Whereas the use of spectral attention results in better performance for some attacks, temporal attention works better for others, and vice versa.
These observations imply that different attacks exhibit different artefacts, none of which can be captured with a single classifier on its own.

\subsection{Fusion}

We performed fusion experiments using the support vector machine (SVM) based fusion approach described in~\cite{IS2020LFCC} using different combinations of LFCC-GMM, GAT-T, GAT-S and RawNet2 systems.  Since the GAT-T and GAT-S systems outperform ResNet-18 systems and since the GAT systems are in any case built on top of the ResNet-18 systems, we discounted the latter in our fusion experiments. Fusion results are presented in boldface in Table~\ref{Tab:comparsion results}. Also included in Table~\ref{Tab:comparsion results} are results for a selection of top-performing, primary systems reported in the literature. These results further demonstrate the effectiveness of GNN-based attention approaches. Fusion results for different combinations of LFCC-GMM, RawNet2 and GAT-based systems all lead to improvements in performance. The best result comes from the fusion of all four systems which represents a 47\% relative reduction in terms of min t-DCF over the best single, GAT-T system. Our results are also competitive with those of competing, top-performing systems. Only one achieves a lower min t-DCF than our best fused system.  Even then, the gap is modest, while our system achieves a lower EER corresponding to a relative improvement of over 10\%.

\begin{table}[!t]
\def\arraystretch{1.1}
	\centering
	\small

	\caption{A performance comparison for the evaluation partition of the ASVspoof 2019 logical access (LA) database in terms of pooled min t-DCF and pooled EER.  Results shown for a set of state-of-the-art countermeasures reported in the literature (regular, roman font) and all systems reported in this paper (illustrated in italics).  Also shown are results for different fusions of the high-spectral resolution LFCC-GMM baseline~\cite{IS2020LFCC}, RawNet2 system and the two GAT-based systems (illustrated in boldface).}
 \setlength\tabcolsep{2pt}
    \begin{tabularx}{\linewidth}{lYY}
		\hline
		 System & min-tDCF& EER	\\ 
	\hline	\hline
	   Spec+LFCC+CQT+SE-Res2Net~\cite{li2020replay}&0.0452&1.89\\
	   \hline
	   {\bfseries LFCC-GMM+GAT-S+GAT-T+RawNet2}&{\bfseries 0.0476}& {\bfseries 1.68}\\
	  \hline
	   LFCC+LFCC-CMVN+CQT+FFT+LCNN+&\multirow{2}{*}{0.0510}&\multirow{2}{*}{1.86}\\
	   LFCC-GMM~\cite{lavrentyeva2019stc}&&\\
	\hline
	ResNet18+LMCL+FM~\cite{chen2020generalization}&0.0520&1.81\\
	\hline
	{\bfseries GAT-S+GAT-T+RawNet2}&{\bfseries 0.0635}& {\bfseries 2.21}\\
	\hline
	{\bfseries LFCC-GMM+RawNet2}&{\bfseries 0.0643}& {\bfseries 2.33}\\
	\hline
	{\bfseries GAT-S+RawNet2}&{\bfseries 0.0692}& {\bfseries 2.29}\\
	\hline
	  Ensemble model~\cite{chettri2019ensemble}&0.0755&2.64\\
		\hline
	{\bfseries GAT-S+GAT-T}&{\bfseries 0.0844}&{\bfseries 4.30}\\
	
	\hline
	{\bfseries GAT-T+RawNet2}&{\bfseries 0.0854}&{ \bfseries 2.61}\\
	\hline
	\emph{GAT-T}&0.0894&4.71\\
	\hline

		 \emph{LFCC-GMM}~\cite{IS2020LFCC} &0.0904&3.50\\
		 \hline
		 \emph{GAT-S}&0.0914&4.48\\
	\hline
	Siamese CNN~\cite{lei2020siamese} &0.0930&3.79\\
	\hline
	FG-CQT+LCNN+CE~\cite{wu2020light}&0.1020&4.07\\
		 \hline
		 LFB-ResNet18~\cite{chen2020generalization}& 0.1090 &4.04\\ 
		 \hline
		 \emph{ResNet-SP}&0.1140&6.82\\
		 \hline
		 \emph{ResNet-ASP}&0.1269&6.22\\
		 \hline
		 \emph{ResNet-SAP}&0.1377&7.11\\
		 \hline
		 
		 \emph{RawNet2}&0.1547&5.54\\
	\hline
		
	
	\end{tabularx}
	\label{Tab:comparsion results}
	\vspace{-0.15cm}
\end{table}

\section{Conclusions}

Graph attention networks (GATs) apply a self-attention mechanism to graph convolutional networks in order to model graph structured data.  Each node in the graph is weighted according to its relevance to other nodes. The weights reflect the relationships between connected nodes, which here represents either a specific sub-band or temporal segment. Our work shows how GATs can be used to model these relationships using high-level representations extracted from deep residual networks and how this improves spoofing detection performance. Our experiments, performed on the ASVspoof 2019 Logical Access dataset, show that the GAT solution outperforms ResNet-18 and RawNet2 baseline systems by a substantial margin and that the GAT system with temporal attention also outperforms the high resolution LFCC-GMM system.  GAT-based systems also outperform all other systems for 9 out of 13 spoofing attacks.  Fusion experiments show that GAT-based systems are complementary to the baselines, with an ensemble system producing a 47\% relative reduction in the t-DCF over the best, single GAT-T system.

While the RawNet2 baseline operates directly upon the raw signal, our GAT solution operates upon filterbank outputs.  We are now working to improve computational efficiency such that our GAT solution can also be applied directly to the raw signal. Another target for our future work will be to determine the nature and origins of the artefacts being detected with spectral and temporal attention and then to link these to specific spoofing attacks and the algorithmic origins.

\section{Acknowledgements}

This work is partly supported by the ExTENSoR project funded by the
French Agence Nationale de la Recherche (ANR) and the VoicePersonae
project funded by ANR and the Japan Science and Technology Agency.

\bibliographystyle{IEEEtran}
\balance
\clearpage
\balance
\bibliography{mybib,references_jung_mendeley}
\end{document}